\documentclass[aps,prd,preprint,tightenlines,superscriptaddress,
amsfonts,amssymb,amsmath,showpacs,nofootinbib]{revtex4-1}
\usepackage[polish,english]{babel}
\usepackage[applemac]{inputenc}
\usepackage[T1]{fontenc}
\usepackage{latexsym}
\usepackage{graphicx}
\usepackage{geometry}
\usepackage{epstopdf}
\usepackage{subfig}
\usepackage{color}
\usepackage[normalem]{ulem}

\usepackage{xcolor}
\usepackage{pdfsync}
\usepackage{fancyhdr}
\usepackage{makeidx}
\usepackage[breaklinks,colorlinks,backref]{hyperref}
\hypersetup{
    colorlinks, %set true if you want colored links
    linktoc=all, %set to all if you want both sections and subsections linked
    linkcolor=red,  %choose some color if you want links to stand out
  citecolor=hyptxt,
  urlcolor=blue}
  \hypersetup{
  citebordercolor=,
  filebordercolor=green,
  linkbordercolor=blue
}
\definecolor{hyptxt}{rgb}{0.7, 0.4, 0.9}
\usepackage{bibtopic}
\newcommand{\bea}{\begin{eqnarray}}
\newcommand{\eea}{\end{eqnarray}}

\newcommand{\dR}{\mathbb R}

\newcommand{\id}{\mathbb I}

\newcommand{\be}{\begin{equation}}
\newcommand{\ee}{\end{equation}}
\newcommand{\I}{\mathbb I}

\newcommand{\ket}[1]{|\kern.3ex#1\kern.3ex\rangle}
\newcommand{\bra}[1]{\langle\kern.3ex #1 \kern.3ex|}
\newcommand{\scalar}[2]{\langle\kern.3ex #1 \kern.3ex|\kern.3ex#2\kern.3ex\rangle}
\newcommand{\norm}[1]{\|\kern.3ex#1\kern.3ex \|}

\newcommand{\RNumb}{\mathbb{R}}

\newcommand{\UnitOp}{\mathbb{I}} % operator unit
\newcommand{\MatUnit}{1\kern-3pt 1} % matrix unit
\newcommand{\Group}[1]{\textrm{#1}} % group assignment
\newcommand{\Bra}[1]{\langle #1 \vert} % bra
\newcommand{\Ket}[1]{\vert #1 \rangle} % ket
\newcommand{\BraKet}[2]{\langle #1 \vert #2 \rangle} % bra(c)ket

\begin{document}

\title{Dependence of the affine coherent states quantization \\
on the parametrization of the affine group}

\author{Andrzej G\'{o}\'{z}d\'{z}}
\email{andrzej.gozdz@umcs.lublin.pl}
\affiliation{orcid:0000-0003-4489-5136; Institute of Physics, Maria Curie-Sk{\l}odowska
University, pl.  Marii Curie-Sk{\l}odowskiej 1, 20-031 Lublin, Poland}

\author{W{\l}odzimierz Piechocki} \email{wlodzimierz.piechocki@ncbj.gov.pl}
\affiliation{orcid:0000-0002-0313-5395; Department of Fundamental Research, National Centre for Nuclear
  Research, Pasteura 7, 02-093 Warszawa, Poland}

\author{Tim Schmitz} \email{tschmitz@thp.uni-koeln.de}\affiliation{orcid:0000-0002-2822-2794; Institut
  f\"{u}r Theoretische Physik, Universit\"{a}t zu K\"{o}ln, Z\"{u}lpicher Stra{\ss}e
  77, 50937 K\"{o}ln, Germany}

%%%%%%%%%%%%%%%%%%%%%%%%%%%%%%%%%%%%%%%%%%%%%%%%%%%%%%%%%%%%%%%%%%%%%%%

\date{\today}

\begin{abstract}
The affine coherent states quantization is a promising integral quantization of Hamiltonian systems
when the phase space includes at least one conjugate pair of variables which takes values from a half-plane.
Such a situation is common for gravitational systems which include singularities. The construction of the
quantization map includes a one-to-one mapping of the half-plane onto the affine group. Particular cases
of this mapping define specific parametrizations of the group. Our aim is showing that different such
parametrizations lead  to unitarily inequivalent quantum theories. Depending on the Hamiltonian system
under consideration, this dependence could potentially be used constructively.\\

\end{abstract}


%\pacs{03.65.--w (Quantum mechanics)}

%%%%%%%%%%%%%%%%%%%%%%%%%%%%%%%%%%%%%%%%%%%%%%%%%%%%%%%%%%%%%%%%%%%
\maketitle
%%%%%%%%%%%%%%%%%%%%%%%%%%%%%%%%%%%%%%%%%%%%%%%%%%%%%%%%%%%%%%%%%%%

%\tableofcontents

%%%%%%%%%%%%%%%%%%%%%%%%%%%%%%%%%%%%%%%%%%%%%%%%%%%%%%%%%%%%%%%%%%%%%%%%
\section{Introduction}
%%%%%%%%%%%%%%%%%%%%%%%%%%%%%%%%%%%%%%%%%%%%%%%%%%%%%%%%%%%%%%%%%%%%%%%%%

The coherent states  quantization can be applied to the quantization of
Hamiltonian systems when the physical phase space can be identified with a Lie group
acting on itself. This group is expected to have a unitary irreducible representation
in a Hilbert space. The latter allows one to construct a resolution of the identity in
that Hilbert space, which can be used to map the observables of that system
into  Hermitian operators.
This quantization method is especially useful in cases where the physical phase
space includes at least one variable with a nontrivial topology, e.g. $\dR_+ =
\{x \in \dR ~|~x >0 \}$, as is common to gravitational systems.

The affine coherent states (ACS) quantization is a special case of the affine
integral quantization, where `affine' refers to the symmetry group of the half-plane
consisting of translations and dilations. Mathematical aspects of this quantization are
presented, for instance,  in the review article \cite{CA}. ACS have been
used in quantizations of numerous physical systems. For a comprehensive review
of  known applications we recommend the references  \cite{JPG} and \cite{Klauder}.
The proceedings of the conference on coherent states   \cite{Proc} present
recent developments.

The ACS quantization of the so-called mixmaster solution of the Einstein equations, examined by
Misner \cite{Mis1,Mis2}, is widely discussed in \cite{Ewa}. It concerns the quantum dynamics of the diagonal
Bianchi IX model. This quantization method, within the semi-classical approximation, leads to the avoidance of
the cosmological singularity. Recent results on quantization of the so-called Belinski-Khalatnikov-Lifshitz
(BKL) scenario \cite{bkl1,bkl2} are much more general \cite{AWG,AWG2}. The classical BKL scenario   was derived by
considering the dynamics of  the non-diagonal (general) Bianchi VIII and IX models, and describes the asymptotic approach to a generic spacelike
singularity in general relativity. The quantum BKL scenario obtained within exact ACS quantization shows that the
gravitational singularity can be replaced by a quantum bounce as a consequence of unitary evolution
of the system, and suggests that quantum general relativity is free from singularities.

Quantization methods usually come with quantization ambiguities. They are generally undesirable
as they lower the predictive power of the resulting quantum theory. The most well known example is probably the freedom to choose different
factor orderings in the usual canonical quantization procedure. In the ACS quantization there is
no factor ordering ambiguity, but one can choose with some freedom the so-called fiducial vector (to be explained later).
Quantization ambiguities  need to be fixed by conceptual criteria and experimental data if available.
On the other hand, the ambiguities are what allow us to tailor the constructed quantum theory to satisfy
these restrictions.
%This might not always be possible in a quantization procedure with less ambiguities.
%For instance, the Groenewold-Van Hove theorems (see, e.g.\ \cite{Gotay} and references therein) says that
%it is impossible to find a quantization map that reproduces the complete algebra of observables.
%One needs to choose a subalgebra that one wants to be reproduced, and quantization ambiguities allow
%to make this choice.

Broad classes of quantization schemes suffer from ambiguities and mathematical inconsistences
(see, e.g.\ \cite{Gotay} and references therein). Since we have growing evidence that the ACS quantization
is capable of dealing with singular gravitational systems, it is reasonable to identify its limitations.

In this article we wish to discuss a quantization ambiguity of the ACS quantization which, as far as
we know, has not been recognized  in the literature before. One may expect that the result of quantization
does not depend on how exactly one identifies the phase space with the Lie group, or in other words
how one parameterizes the group, but this is not the case. In fact, different parametrizations
lead to quantum systems which are not unitarily equivalent, i.e. represent different quantum systems.
The aim of this paper is showing explicitly this intriguing property of the ACS quantization.

In what follows, we first present two parametrizations of the affine
group used in the literature and derive the two corresponding quantizations
which turn out to be unitarily inequivalent. Next, we extend this result to
the general case. Finally, we conclude.

%%%%%%%%%%%%%%%%%%%%%%%%%%%%%%%%%%%%%%%%%%%%%%%%%%%%%%%%%%%%%%%%%%
\section{Comparing two known parametrizations} \label{known}
%%%%%%%%%%%%%%%%%%%%%%%%%%%%%%%%%%%%%%%%%%%%%%%%%%%%%%%%%%%%%%%%%%%%

To make the present paper self-contained we first recall general ideas
underlying coherent states (see, e.g. \cite{Perel}).
For a Lie group $G$, let $U(g)$, where $g\in G$, be a unitary irreducible
representation of it in some Hilbert space $\mathcal{H}$.  One can take an (at
this point) arbitrary $\ket{\Phi}\in\mathcal{H}$, called fiducial vector, and
act on it with $U(g)$ as follows
\begin{equation}\label{csq1}
	\ket{g}=U(g)\ket{\Phi}\,,
\end{equation}
to construct a family of coherent states.

 Consider the operator
\begin{equation}\label{csq2}
	\mathcal{O}=\int_G d\mu(g)~\ket{g}\bra{g}=
\int_G d\mu(g)~U(g)\ket{\Phi}\bra{\Phi}U^\dagger(g)\,,
\end{equation}
where $d\mu(g_0\cdot g)=d\mu(g)$ is a left invariant measure on $G$. It is easy
to see that $\mathcal{O}$ intertwines $U(g)$,
\begin{align}\label{csq3}
	U(g)\cdot \mathcal{O} =& \int_G d\mu(g')~U(g\cdot
        g')\ket{\Phi}\bra{\Phi}U(g'^{-1}) \nonumber\\
\overset{h=g\cdot g'}{=}
	& \int_G d\mu(h)~U(h)\ket{\Phi}\bra{\Phi}U(h^{-1}\cdot g)=
\mathcal{O}\cdot U(g)\,.
\end{align}
As we know from Schur's Lemma, any non-trivial intertwiner is a scalar multiple
of the identity, i.e.  $\mathcal{O}\propto \I_\mathcal{H}$.  The factor of
proportionality has to be decided on a case by case basis, and may come with a
restriction on the fiducial vector $\ket{\Phi}$.

The above means that if $U$ is the unitary irreducible representation of $G$ in the Hilbert space
$\mathcal{H}$, the family of coherent states \eqref{csq1} can be used to define a resolution of the identity in $\mathcal{H}$. The latter is of primary importance in the quantization procedure described below.

In what follows we discuss coherent states constructed as above from the affine
group, affine coherent states (ACS), and show how they can be used in
quantization.  This procedure is called the affine coherent state quantization,
and we give here a short introduction to this method.

In this section we compare the ACS quantization corresponding to two simple parametrizations of the affine group.
The general case is considered in the next section.

Suppose the phase space of some physical system is a half plane, $\Pi = \{
(p,q)\in \dR \times \dR_+ \}$. It can be identified with the affine group $G :=$
Aff$(\dR)$ by defining the multiplication law either by (see \cite{AWG} for more
details)
\begin{equation}\label{m1}
  (p_1, q_1)_1\cdot (p_2, q_2)_1 := (q_1 p_2 + p_1, q_1 q_2)_1\, ,
\end{equation}
or by (see \cite{CA} for more details)
\begin{equation}\label{m2}
  (\tilde{p}_1,\tilde{q}_1 )_2\cdot (\tilde{p}_2,\tilde{q}_2)_2 :=
(\tilde{p}_2/ \tilde{q}_1 + \tilde{p}_1,  \tilde{q}_1 \tilde{q}_2)_2\, .
\end{equation}
Eqs. \eqref{m1}--\eqref{m2} define two different parametrizations\footnote{To make possible the comparison
of the parametrizations used in \cite{AWG} and \cite{CA}, we rename the variables $ (p,q) \in \dR_+ \times \dR $ of \cite{AWG}
so that $(p,q) \in \dR \times \dR_+$,  in the present paper, which fits the notation of \cite{CA}.} of $G$.
They correspond, respectively, to the two actions of this group on $\dR_+$:
\begin{equation}\label{m3}
  x^\prime = (p,q)_1 \cdot x := x q +p~~~\mbox{and}~~~x^\prime =
(\tilde{p},\tilde{q})_2 \cdot x := x/\tilde{q} + \tilde{p} \, .
\end{equation}

The affine group has two (nontrivial) inequivalent irreducible unitary
representations, \cite{Gel} and \cite{AK1,AK2}, defined in the Hilbert space\footnote{The representation
defined in \cite{CA} is $U(p,q) \psi(x) = \frac{e^{ipx}}{\sqrt{q}}\psi(x/q)$ with the carrier
space $L^2(\dR_+, dx)$, but takes the form $ U(p,q) \psi(x) = e^{ipx}\psi(x/q)$ when acting in
$L^2(\dR_+, dx/x)$. Since the measure $dx/x$ is invariant for dilations on $\dR_+$, it is natural to use
the latter space as the carrier space of the affine group representation. Similarly, $L^2(\dR, dx)$ is the natural
carrier space for the representation of the additive group on $\dR$ because $dx$ is invariant on $\dR$.}
$\mathcal{H}:= L^2 (\dR_+, d\nu(x))$, where $d\nu(x):= dx/x$. For both
parametrizations we choose the one defined, respectively, by
\begin{equation}\label{m4}
  U_1 (p,q)\Psi (x) := e^{ipx} \Psi (qx)~~~\mbox{and}~~~
U_2 (\tilde{p},\tilde{q})\Psi (x) := e^{i\tilde{p}x} \Psi (x/\tilde{q}) \, ,
\end{equation}
where $\Psi (x) = \langle x|\Psi \rangle$ and $|\Psi\rangle\in \mathcal{H}$.

Integration over the affine group is defined, respectively, as
\begin{equation}\label{m5}
  \int_G d\mu_1 (p,q) := \int_{-\infty}^\infty dp \int_0^\infty dq /q^2~~~
\mbox{and}~~~
  \int_G d\mu_2 (\tilde{p},\tilde{q}):=
\int_{-\infty}^\infty d\tilde{p} \int_0^\infty d\tilde{q} \, ,
\end{equation}
where both measures in \eqref{m5} are left invariant.

Any coherent state can be obtained as
\begin{equation}\label{m6}
  \langle x |p,q \rangle _1= U_1 (p,q)\Phi (x)~~~
\mbox{or}~~~\langle x |\tilde{p},\tilde{q} \rangle _2=
U_2 (\tilde{p},\tilde{q})\Phi (x),
\end{equation}
where the fiducial vector $\Phi (x) = \langle x | \Phi \rangle$, $|\Phi \rangle \in \mathcal{H}$,
is required to satisfy  $\langle \Phi|\Phi \rangle =1$.

The resolutions of the identity in the Hilbert space $\mathcal{H}$
read
\begin{equation}\label{m7}
  \int_G d\mu_1 (p,q)| p,q\rangle_1 ~ {_1} \langle p,q | = 2\pi
A_\Phi\I~~~\mbox{and}~~~
  \int_G d\mu_2 (\tilde{p},\tilde{q})| \tilde{p},
\tilde{q}\rangle_2 ~ {_2} \langle \tilde{p},\tilde{q} | = 2\pi A_\Phi \I \, ,
\end{equation}
where
\begin{equation}\label{m8}
 A_\Phi = \int_0^\infty \frac{dx}{x^2} |\Phi(x)|^2  < \infty \, .
\end{equation}
Eq. \!\eqref{m8} defines an additional condition we impose on the fiducial vector $\Phi(x)$.

Making use of \eqref{m7} one can (formally) map any observable $f: \Pi \rightarrow \dR$
into a symmetric operator  $\hat{f}: \mathcal{H}\rightarrow \mathcal{H}$ as
follows:
\begin{equation}\label{m9}
  \hat{f}_1 := \frac{1}{2\pi A_\Phi}\int_G d\mu_1 (p,q)| p,q\rangle_1~f(p,q)~ {_1}
\langle p,q |,
\end{equation}
or
\begin{equation}\label{mm9}
\hat{f}_2 := \frac{1}{2\pi A_\Phi}\int_G d\mu_2 (\tilde{p},\tilde{q})|
\tilde{p},\tilde{q}\rangle_2~f(\tilde{p},\tilde{q})~{_2} \langle \tilde{p},\tilde{q} | \, .
\end{equation}

Due to the above we have
\begin{equation}\label{par1}
 \hat{f}_1 \Psi (x) = \frac{1}{2\pi A_\Phi}\int_G d\mu_1(p,q)
e^{ipx}\Phi (qx) f(p,q)~{_1} \langle p,q |\Psi\rangle \, ,
\end{equation}
where
\begin{equation}\label{par2}
  {_1} \langle p,q |\Psi\rangle =
\int_0^\infty d\nu (x^\prime)
e^{-ipx^\prime}\Phi(qx^\prime)^\ast \Psi (x^\prime) \, ,
\end{equation}
and
\begin{equation}\label{par3}
 \hat{f}_2 \Psi (x) = \frac{1}{2\pi A_\Phi}\int_G d\mu_2(\tilde{p},\tilde{q})
e^{i\tilde{p}x}\Phi (x/\tilde{q})
f(\tilde{p},\tilde{q})~{_2} \langle \tilde{p},\tilde{q}|\Psi\rangle \, ,
\end{equation}
where
\begin{equation}\label{par4}
  {_2} \langle \tilde{p},\tilde{q} |\Psi\rangle =
\int_0^\infty d\nu (x^\prime)
e^{-i\tilde{p}x^\prime}\Phi(x^\prime/\tilde{q})^\ast \Psi (x^\prime) \, .
\end{equation}

To proceed in calculations we use in what follows the following identity defined in $\mathcal{H}$
and derived in App.\ \ref{aaa}:
\begin{equation}\label{dist}
\id = \int_0^\infty
d \nu (x) | x \rangle \langle x| \, .
\end{equation}

To get \eqref{par2} and \eqref{par4}, we apply  \eqref{dist}  to each of the left hand side of these equations,
and Eq. \!\eqref{m6} together with Eq. \!\eqref{m4}.
Making use of the substitution $q = 1/\tilde{q}~$ in Eq. \!\eqref{par1},  we easily obtain
\begin{equation}\label{par6}
 \hat{f}_1 \Psi (x) =
\frac{1}{2\pi A_\Phi}\int_G d\mu_2(\tilde{p},\tilde{q})
e^{i\tilde{p}x}\Phi (x/\tilde{q}) f(\tilde{p},1/ \tilde{q})
 ~{_2} \langle \tilde{p},\tilde{q}|\Psi\rangle \, .
\end{equation}
Comparing \eqref{par3} with \eqref{par6} we can see that for a generic $|\Psi\rangle \in
\mathcal{H}$ we have $ \hat{f}_2 |\Psi\rangle \neq \hat{f}_1 |\Psi\rangle$, as in general
$f(\tilde{p}, \tilde{q}) \neq f(\tilde{p}, 1/\tilde{q}) $. It means that these operators
act quite differently in $\mathcal{H}$.

Are operators, however, unitarily equivalent?
To answer this essential question, we compare the traces of both operators in some orthonormal
basis $\{|e_k \rangle \}$ of $\mathcal{H}$ to see\footnote{The trace of an operator does not depend
on the choice of basis.} whether
\begin{equation}\label{tr1}
  \mathrm{Tr} (\hat{f}_1) =  \mathrm{Tr} (\hat{f}_2) \, .
\end{equation}
Eq. \eqref{tr1} is satisfied if the operators $\hat{f}_1$ and $\hat{f}_2$ are
unitarily equivalent, since  they have the same trace:
\begin{equation}\label{ttr}
\mathrm{Tr}(\check{U}\hat{f}{\check{U}}^{-1}) = \mathrm{Tr}(\hat{f}\check{U}{\check{U}}^{-1}) = \mathrm{Tr}(\hat{f}),
\end{equation}
where $\check{U}$ is some unitary operator.

The main part of the above verification is the rewriting:
\begin{align}\nonumber
 \mathrm{Tr} (\hat{f}_1) = \sum_n \langle e_n| \hat{f}_1 | e_n \rangle &=
\frac{1}{2\pi A_\Phi}
\sum_n \int_G d\mu_1(p,q)\,\langle e_n| p, q \rangle_1
~f(p,q)~{_1} \langle p, q |e_n \rangle\\
\nonumber
 &= \frac{1}{2\pi A_\Phi}
\sum_n \int_G d\mu_2 (\tilde{p},\tilde{q})\,\langle e_n| \tilde{p},
 \tilde{q}\rangle_2
~f(\tilde{p},1/\tilde{q})~{_2} \langle \tilde{p},\tilde{q}|e_n \rangle\\
\nonumber
  &=\frac{1}{2\pi A_\Phi} \int_G d\mu_2 (\tilde{p},\tilde{q})\,~{_2} \langle \tilde{p},\tilde{q}|
\sum_n|e_n \rangle \langle e_n| |
\tilde{p}, \tilde{q}\rangle_2 ~f(\tilde{p},1/\tilde{q})\\\nonumber
  &=\frac{1}{2\pi A_\Phi} \int_G d\mu_2 (\tilde{p},\tilde{q}) \,~{_2} \langle \tilde{p},\tilde{q}|
\tilde{p}, \tilde{q}\rangle_2 ~f(\tilde{p},1/\tilde{q})\\
\label{tr2}
     &=\frac{1}{2\pi A_\Phi} \int_G d\mu_2 (\tilde{p},\tilde{q})\,f(\tilde{p},1/\tilde{q}) \, .
 \end{align}
The transition from one parametrization to another, marked by the third equality sign in Eq. \!\eqref{tr2},
can be obtained in a similar way as the transition from  Eqs. \!\eqref{par1}--\eqref{par2} to Eq. \!\eqref{par6}.
The derivation of \eqref{tr2} makes use of the equations
\begin{equation}\label{tr3}
  \sum_n|e_n \rangle \langle e_n|  =
\I~~~\mbox{and}~~~{_2} \langle \tilde{p},\tilde{q}|
  \tilde{p} \tilde{q}\rangle_2 =
\langle \Phi|U^{-1}_2 U^{\,}_2 |\Phi \rangle = \langle\Phi | \Phi \rangle = 1\, .
 \end{equation}
 Similarly, we get
 \begin{equation}\label{tr4}
 \mathrm{Tr} (\hat{f}_2) = \frac{1}{2\pi A_\Phi}\int_G d\mu_2 (\tilde{p},\tilde{q})\,f(\tilde{p},\tilde{q}) \, .
 \end{equation}

Therefore, Eq. \eqref{tr1} cannot be satisfied, as $f(\tilde{p}, \tilde{q}) \neq f(\tilde{p}, 1/\tilde{q})$, so that
the considered operators, $\hat{f_1}$ and $\hat{f_2}$,   are unitarily inequivalent. It further means that the two
considered affine group parametrizations, defined by \eqref{m1}--\eqref{m2}, lead to quite different quantum systems.

 \section{Considering the general case}

 In this section we consider an arbitrary  parametrization of the affine group.
Different parametrizations of the affine group can  be implemented by a family of one-to-one transformations
$\chi: \Pi \to \Group{Aff(R)}$. Every function $\chi$ provides a parametrization of the affine group by elements
of the phase space $\Pi$ as follows
\begin{equation}
\label{ParamPhaceSpGrp}
\chi(p,q)=(\xi(p,q),\eta(p,q)) \in \Group{Aff(R)}.
\end{equation}
To define a general parametrization, one can use as an intermediate step either \eqref{m1} or \eqref{m2}.
To be specific, we choose \eqref{m1} so that we have the composition law in the form
\begin{multline}
\label{MultiplicLaw2}
(\xi(p_1,q_1),\eta(p_1,q_1)) \cdot (\xi(p_2,q_2),\eta(p_2,q_2))\\=
((\eta(p_1,q_1)\xi(p_2,q_2) +\xi(p_1,q_1),\eta(p_1,q_1)\eta(p_2,q_2)) \, .
\end{multline}
This determines uniquely the composition law for the new parametrization of the affine group.

The corresponding invariant measure can be obtained by the change of variables in
the first measure in (\ref{m5}):
\begin{equation}
\label{GeneralInvMeasure}
d\xi \, \frac{d\eta}{\eta^2}=
\left[\frac{1}{\eta(p,q)}\right]^2
\left| \frac{\partial(\xi,\eta)}{\partial(p,q)} \right| dp \, dq =:
\sigma(p,q) \, dp \, dq \, .
\end{equation}
Therefore, the ACS quantization of the phase space function $f: \Pi \rightarrow \dR$ yields
\begin{align}\nonumber
 \hat{f}  &=\frac{1}{2\pi A_\Phi}\int_{G}  \,d\xi \, \frac{d\eta}{\eta^2} \,
\Ket{\xi,\eta} f(p(\xi,\eta),q(\xi,\eta)) \Bra{\xi,\eta}\\
\label{GeneralACSQ}
     &= \frac{1}{2\pi A_\Phi}\int_{G}  \, dp \, dq \,  \sigma(p,q)
\Ket{\xi(p,q),\eta(p,q)} f(p,q) \Bra{\xi(p,q),\eta(p,q)} \, .
 \end{align}
Following the idea of the preceding section, we calculate the trace of  $\hat{f}$.
If the trace of $\hat{f}$ is independent on the affine group
parametrization, then the ACS method of quantization is universal. But we have
\begin{eqnarray}
\label{TraceOpf}
 \mathrm{Tr}(\hat{f})&&=\frac{1}{2\pi A_\Phi}
\int_{G}\, dp \, dq\,  \sigma(p,q) f(p,q)
\mathrm{Tr}(\Ket{\xi(p,q),\eta(p,q)} \Bra{\xi(p,q),\eta(p,q)}) \nonumber \\
&& = \frac{1}{2\pi A_\Phi}\int_{G}\, dp \, dq\,  \sigma(p,q) f(p,q).
\end{eqnarray}
To obtain the second equality in the above equation we have used \eqref{tr3}.

Eq. \!\eqref{TraceOpf} shows explicitly  the dependence of the trace of the operator $\hat{f}$ on the
parametrization \eqref{ParamPhaceSpGrp} due to the term $\sigma(p,q)$ in the integrant. Therefore,
the ACS quantization scheme depends on the parametrization of the group manifold by the phase space
variables. The implications of this dependence are discussed in the conclusions.

\section{Conclusions}

The result we have obtained is quite general. The dependence we have found can be seen as an advantage
of this method over other quantization schemes as it allows one to construct quantum theories fulfilling specific
requirements. However, this high flexibility may lead to low predictability of the resulting quantum theory. Therefore, an additional constraint
on the choice of the parametrization, well motivated physically,  should be an essential element of the
constructed integral quantization scheme.

In the case of quantization of the BKL scenario \cite{AWG} we have found \cite{AWG2} that the two
simplest parametrizations of the affine group, considered in Sec. \!\ref{known}, lead to qualitatively
the same results. The difference concerns quantitative details not essential to the main conclusion that a
regular quantum bounce replaces the classical generic singularity of the BKL scenario. In this case,
the conceptual criterion of having regular quantum dynamics does not single out a parametrization, neither is this necessary. However, obtaining this result was made possible by taking a simplified
form of the classical Hamiltonian that describes the dynamics properly only in the close vicinity of the
gravitational singularity. Considering the exact Hamiltonian would probably lead to making use of the freedom
in the choice of the group parametrization  to meet this conceptual criterion.

In App.\ \ref{bbb} we consider the issue of quantization of the affine group algebra. It is shown
that the parametrization of \cite{AWG} fails in reproducing the classical algebra, whereas the one
of  \cite{CA} is successful. This is a further possible criterion for choosing a particular parametrization.

The issue of mapping a classical observable $f: \Pi \rightarrow \dR$ into a self-adjoint operator
$\hat{f}: \mathcal{H} \rightarrow  \mathcal{H}\,$ is of basic importance in any quantization
scheme\footnote{In fact, only a self-adjoint operator can represent an observable at the quantum level.}.
It is highly problematic if $\hat{f}$ is an unbounded operator (see, e.g. \cite{Reed,Kon}). In the case
of the ACS quantization, the dependence on the group parametrization can be helpful. To be specific,
let us consider the norm of $\hat{f}$ defined by \eqref{GeneralACSQ}):
\begin{align}
\label{GeneralACSQHam}
\Vert\hat{f} \Vert & \leq
\frac{1}{2\pi A_\Phi}\int_{G}  \, dp \, dq \, \Vert \sigma(p,q)
\Ket{\xi(p,q),\eta(p,q)} f(p,q) \Bra{\xi(p,q),\eta(p,q)} \Vert \nonumber \\
& = \frac{1}{2\pi A_\Phi}\int_{G}  \, dp \, dq \, |\sigma(p,q) f(p,q)|\,
  \Vert \Ket{\xi(p,q),\eta(p,q)} \Bra{\xi(p,qk),\eta(p,q)} \Vert \nonumber \\
& = \frac{1}{2\pi A_\Phi}\int_{G}  \, dp \, dq \, |\sigma(p,q) f(p,q)|  \, ,
 \end{align}
 where the projection operator
 $\Ket{\xi(p,q),\eta(p,q)} \Bra{\xi(p,qk),\eta(p,q)}$ has the norm \\
  $\Vert \, \Ket{\xi(p,q),\eta(p,q)} \Bra{\xi(p,qk),\eta(p,q)} \, \Vert =1$.

 The inequality (\ref{GeneralACSQHam}) shows that the symmetric operator $\hat{f}$ is bounded so that self-adjoint if
\begin{equation}
 \label{GeneralACSQHam2}
\frac{1}{2\pi A_\Phi}\int_{G}  \, dp \, dq \,
|\sigma(p,q) f(p,q)| < \infty  \, .
 \end{equation}
 The open question is whether one can  always find a parametrization of the group
 manifold which, due to the form of the Jacobi determinant $\sigma(p,q)$,
 fulfills the condition (\ref{GeneralACSQHam2}). Each case needs a separate examination.

 In summary, one can say that the freedom in the choice of the group parametrization in the ACS quantization
 is a free `parameter' of this quantization scheme. However, one should be conscious
 that different parametrizations lead to unitarily inequivalent quantum systems.

In our recent paper \cite{WT} we have successfully  used  the above freedom to interpret the connection between
quantum theories for the comoving observer and the stationary exterior observer of the Oppenheimer-Snyder collapse model.
This issue was difficult to treat within the canonical quantization approach \cite{Schmitz:2019jct}. The freedom
in the choice of the group parametrization may turn out to be useful in the ACS quantization of the Lema\^{i}tre-Tolman-Bondi
model for gravitational collapse  that has already been quantized canonically \cite{Kiefer:2019csi}.
Possible agreement of the results obtained within both quantization schemes may serve to demonstrate the robustness
of the obtained results.

%%%%%%%%%%%%%%%%%%%%%%%%%%%%%%%%%%%%%%%%%%%%%%%%%%%%%%%%%%%%%%%%%%%%%%%%%%%%%%%%%%%%%%%%%%%%%%%%%%%%%%%%%
\acknowledgments This work was partially supported by the German-Polish
bilateral project DAAD and MNiSW, No 57391638.
%%%%%%%%%%%%%%%%%%%%%%%%%%%%%%%%%%%%%%%%%%%%%%%%%%%%%%%%%%%%%%%%%%%%%%%%%%%%%%%%%%%%%%%%%%%%%%%%%%%%%%%%%

\appendix

\section{Subsidiary identity}\label{aaa}

Let us consider two spaces of square integrable functions:
$\mathcal{K}_1=\mathrm{L}^2(\RNumb,dy)$ and
$\mathcal{K}_2=\mathrm{L}^2(\RNumb_+,dx/x)$, where $\Group{G}_1=(\RNumb,+)$ is
the additive group of real numbers and $\Group{G}_2=(\RNumb_+,\cdot)$ is the
multiplicative group of positive real numbers.

The measure $dy$ is the invariant measure for $G_1$ and
$d\nu(x)=\frac{dx}{x}$ is the invariant measure for $G_2$.

The logarithmic function gives an isomorphism between both groups:
\begin{equation}
\label{RplusRIsomorph}
\ln: \Group{G}_2 \to \Group{G}_1,  \quad y=\ln(x) \ .
\end{equation}
This means that one can transfer a part of the notions well defined on one group to
the second one. For example, the Dirac delta function $\delta(y)$ defined on
$\RNumb$ determines the Dirac type delta function $\delta_{\RNumb_+}(x)$ defined on $\dR_+$:
\begin{equation}
\label{DiracDeltaG}
\delta_{\RNumb_+}(x)=\delta(\ln(x))=\delta(y), \quad x>0 \ .
\end{equation}
This implies (renaming $\phi(x)=\psi(\ln(x))=\psi(y)$) that
\begin{equation}
\label{DiracDeltaG3}
\int_{\RNumb_+} \phi(x) \delta_{\RNumb_+}(x) d\nu(x) =  \phi(1) \, .
\end{equation}
In both spaces $\mathcal{K}_1$ and $\mathcal{K}_2$  the position operator is
defined as the multiplication operator (in $\mathcal{K}_2$ the additional
constraint $y > 0$ is required)
\begin{equation}
\label{PositionOper}
\hat{x} \psi(y) = y \psi(y) \, .
\end{equation}
The generalized eigenvectors of $\hat{x}$ corresponding to the eigenvalue $x$
are the Dirac delta distribution $\delta(y-x)$ in $\mathcal{K}_1$ and, by making
use of the logarithmic transformation, the distribution $\delta_{\RNumb_+}(x^{-1}y)$
in the space $\mathcal{K}_2$. This and (\ref{DiracDeltaG3}) imply
\begin{equation}
\label{PositionOperRplus7}
\BraKet{x}{\phi}:= \int_{\RNumb_+} d\nu(y) \delta_{\RNumb_+}(x^{-1}y) \phi(y)=\phi(x)\, .
\end{equation}
Using the above property, one gets
\begin{eqnarray}
\label{PositionOperRplus8}
&& \BraKet{\phi_2}{\phi_1} =
\int_{\RNumb_+} d\nu(x) \BraKet{x}{\phi_2}^\star \BraKet{x}{\phi_1}
\nonumber \\
&& = \Bra{\phi_2} \left\{ \int_{\RNumb_+}
d\nu(x) \Ket{x} \Bra{x} \right\} \Ket{\phi_1}
\end{eqnarray}
for all $\phi_1$ and $\phi_2$, which implies:
\begin{equation}
\label{PositionOperRplus9}
\int_{\RNumb_+} d\nu(x) \Ket{x} \Bra{x} = \UnitOp \, .
\end{equation}

\section{Example of choosing a suitable parametrization}\label{bbb}

In this section we  focus on the reproduction of the classical algebra of a subset of observables of the theory.
Usually, the prime example would be the canonical commutation relation $[\hat{q},\hat{p}]=i\hbar$, but on the half
line it is problematic: $\hat{p}$ is not strictly speaking an observable in the quantum theory, since it cannot be
made self-adjoint.

Instead we want to focus on the
classical relation $\{q,d\}=q$, where $d:=pq$, as was done for example in \cite{Klauder}. The quantum counterpart
of this is called the affine commutation relation. The operator $\hat{d}$ generates dilations, which in contrast
to the translations generated by $\hat{p}$ can not push position eigenstates off of the half line. This makes
the affine commutation the natural replacement for the canonical commutation relation for quantum theories on
the half line. In addition, this algebra is also of importance to some approaches to quantum gravity,
see e.g.\ \cite{AffineLQG,AffineKlauder}.

For simplicity we only consider real fiducial vectors, and restrict ourselves to the two specific parametrizations used
in section \ref{known}. A quick calculation shows that the respective position operators are given as
\begin{equation}\label{ex1}
	\hat{q}_1\psi(x)=\frac{1}{A_\Phi}\frac{\psi(x)}{x},~~~
	\mbox{and}~~~\hat{q}_2\psi(x)=\frac{B_\Phi}{A_\Phi}\,x\, \psi(x),
\end{equation}
where
\begin{equation}
	B_\Phi=\int_0^\infty\frac{dx}{x^3}\Phi(x)^2,
\end{equation}
and the dilation operators as
\begin{equation}
	\hat{d}_1\psi(x)=-\frac{i}{A_\Phi}\frac{d}{dx}\left( \frac{\psi(x)}{x}\right) ,~~~
	\mbox{and}~~~\hat{d}_2\psi(x)=-i\frac{B_\Phi}{A_\Phi}\,x\, \frac{d}{dx}\psi(x).
\end{equation}
This leads to
\begin{equation}
	[\hat{q}_1,\hat{d}_1] = i A_\Phi \;\hat{q}_1^3,~~~
	\mbox{and}~~~[\hat{q}_2,\hat{d}_2]=i\frac{B_\Phi}{A_\Phi}\;\hat{q}_2.
\end{equation}

Apart from numerical prefactors depending on $\Phi(x)$, which could be absorbed through a rescaling of $p$,
the parametrization of \cite{CA} fulfills the affine commutation relation, and the one from \cite{AWG} does
not. Based on this criterion, one should hence choose the former one.

%%%%%%%%%%%%%%%%%%%%%%%%%%%%%%%%%%%%%%%%%%%%%%%%%%%%%%%%%%%%%%%%%%%%%%%%%%

\end{document}